\documentclass[12pt]{article}

\def\fun#1#2{\lower3.6pt\vbox{\baselineskip0pt\lineskip.9pt
  \ialign{$\mathsurround=0pt#1\hfil##\hfil$\crcr#2\crcr\sim\crcr}}}

\begin{document}

\title{Gravitational Lensing
in Inhomogeneous Universes} 
\author{Robert M. Wald\\ 
{\it Enrico Fermi Institute and Department of Physics}\\ 
{\it University of Chicago}\\
{\it 5640 S. Ellis Avenue}\\ 
{\it Chicago, Illinois 60637-1433}}
\maketitle

\begin{abstract}

I describe a new approach (developed in collaboration with D.E. Holz)
to calculating the statistical distributions for magnification, shear,
and rotation of images of cosmological sources due to gravitational
lensing by mass inhomogeneities on galactic and smaller scales. Our
approach is somewhat similar to that used in ``Swiss cheese'' models,
but the ``cheese'' has been completely eliminated, the matter
distribution in the ``voids'' need not be spherically symmetric, the
total mass in each void need equal the corresponding Robertson-Walker
mass only on average, and we do not impose an ``opaque radius''
cutoff. In our approach, we integrate the geodesic deviation equation
backwards in time until the desired redshift is reached, using a Monte
Carlo procedure wherein each photon beam in effect ``creates its own
universe'' as it propagates. Our approach fully takes into account
effects of multiple encounters with gravitational lenses and is much
easier to apply than ``ray shooting'' methods.

\end{abstract}

In this paper, I will briefly summarize a new approach to determining
statistical information on the magnification, shear, and rotation of
images of cosmological sources due to gravitational lensing by mass
inhomogeneities on galactic and smaller scales. In this approach, one
may freely specify an underlying Robertson-Walker cosmological model together
with the relevant information on how the matter is distributed in
galaxies. This approach was developed in collaboration with Daniel
E. Holz. Full details of the method are given in \cite{hw} and some
results and applications are given in \cite{hw}, \cite{hmq}, and
\cite{h}.

The main assumption underlying our approach concerns the spacetime
structure of our universe. It is generally believed that the
Robertson-Walker models provide an excellent description of the
spacetime metric of our universe ``on large scales''. However,
Robertson-Walker metrics provide an extremely poor approximation to
the description of the actual matter density and spacetime curvature
on small scales, which, for example, on Earth are a factor of
$~10^{30}$ higher than given by a typical Robertson-Walker model. On
the other hand, apart from negligibly small regions of spacetime in
the immediate proximity of black holes and neutron stars, Newtonian
gravity appears to provide an excellent description of all
gravitational phenomena on scales much smaller than the Hubble
radius. Thus, the spacetime metric of our universe appears to be such
that it corresponds to a Robertson-Walker metric on large scales and a
Newtonian metric (relative to the local Robertson-Walker rest frame)
on small scales.

A spacetime metric with these features is
\begin{equation} 
ds^2 = -(1 + 2\phi)\,d \tau^2 + (1 - 2\phi) a^2(\tau)
\left[\frac{dr^2}{1-kr^2} +
r^2 (d \theta^2 + sin^2 \theta\,d \varphi^2)\right],
\label{metric} 
\end{equation}
provided that
\begin{equation} 
|\phi| \ll 1 .
\label{d1phi} 
\end{equation}
\begin{equation} 
|\partial \phi/ \partial \tau|^2 \ll a^{-2} h^{ab} D_a \phi D_b \phi,
\label{d2phi} 
\end{equation}
\begin{equation} 
(h^{ab} D_a \phi D_b \phi)^2 \ll h^{ac} h^{bd} D_a D_b \phi D_c D_d \phi.
\label{d3phi} 
\end{equation}
and the matter stress-energy tensor, $T_{ab}$
({\em not} including the cosmological constant term) satisfies
\begin{equation} 
T_{ab} \approx \rho u_a u_b,
\label{Tab} 
\end{equation}
where $u^a$ is a unit vector which points in the $(\partial / \partial
\tau)^a$ direction. In the above equations, $h_{ab}$ is either the
metric of a unit 3-sphere ($k = 1$), a unit 3-hyperboloid ($k=-1$), or
flat 3-space ($k = 0$)
\begin{equation} 
h_{ab} \equiv \frac{1}{1-kr^2}\,dr_a dr_b + r^2 (d \theta_a d \theta_b +
sin^2 \theta\,d \varphi_a d \varphi_b)
\label{smetric} 
\end{equation}
and $D_a$ denotes the spatial derivative operator associated with
$h_{ab}$. In essence, eq.(\ref{d1phi}) ensures that the large scale
structure of the universe is well described by an {\em underlying
Robertson-Walker model} (i.e., the metric obtained by setting $\phi =
0$ in eq.(\ref{metric})). Equation (\ref{d2phi}) ensures that the
metric is locally quasi-static in the rest frame of a Robertson-Walker
observer (so that, in particular, negliglible gravitational radiation
is present) and eq.(\ref{d3phi}) ensures that the nonlinear
contributions of $\phi$ to the curvature are negligible compared with
the linear contributions. These latter two conditions are necessary for
Newtonian gravity to be a good approximation locally (i.e., on scales
small compared with the Hubble radius) in the Robertson-Walker rest
frame. Finally eq.(\ref{Tab}) assumes that matter stresses are small
compared with energy densities and that the matter does not move at
speeds comparable to $c$ relative to the rest frame of the underlying
Robertson-Walker model. It is important to note that
eqs.(\ref{d1phi})-(\ref{Tab}) {\it do} permit the spatial derivatives
of $\phi$ to locally be very large compared with scales set by the
curvature of the underlying Robertson-Walker model.

As found in \cite{hw}, the metric (\ref{metric}) and
eqs.(\ref{d1phi})-(\ref{d3phi}) are compatible with Einstein's
equation\footnote{More precisely, if eqs.(\ref{ee1'})-(\ref{poisson})
hold, then all components of Einstein's equation except the
``time-space'' components will be satisfied up to terms which are
negligible compared with the curvature of the underlying
Robertson-Walker model or which can be neglected by
eqs.(\ref{d1phi})-(\ref{d3phi}). As noted in \cite{hw}, the
``time-space'' components of Einstein's equation contains terms which
need not be locally small compared with the Robertson-Walker
curvature, in which case one must generalize the metric (\ref{metric})
so as to allow for the presence of novanishing ``time-space''
components of the metric. However, these components will be small
compared with $\phi$ and will have completely negligible effects on
lensing.}  provided that $a$ is related to the spatially averaged mass
density $\bar{\rho}$ by the usual Robertson-Walker equations
\begin{eqnarray}
&3 \ddot{a}/a = \Lambda - 4 \pi \bar{\rho}&
\label{ee1'} \\
&3(\dot{a}/a)^2 = \Lambda + 8 \pi \bar{\rho} - 3k/a^2,&
\label{ee2'} 
\end{eqnarray}
and $\phi$ satisfies 
\begin{equation} 
a^{-2} h^{ab} D_a D_b \phi = 4 \pi \delta \rho.
\label{poisson} 
\end{equation}
It then follows \cite{hw} that in a local inertial frame associated with
any Robertson-Walker observer, the metric takes the form
\begin{equation} 
ds^2 = - (1 + 2 \Phi - \Lambda R^2/3)\,dT^2 + (1 - 2 \Phi - \Lambda
R^2/6)[dX^2 + dY^2 + dZ^2],
\label{n2} 
\end{equation}
where $R^2 = X^2 + Y^2 + Z^2$ and terms of order $(R/R_H)^3$ and
higher have been dropped, where $R_H \equiv a/\dot{a}$. Here
\begin{equation} 
\Phi \equiv \phi + 2 \pi R^2 \bar{\rho}/3,
\label{Phi} 
\end{equation}
satisfies the ordinary Poisson equation
\begin{equation} 
\nabla^2 \Phi = 4\pi \rho.
\label{Poisson}
\end{equation}
Thus, eq.(\ref{n2}) describes a Newtonian perturbation of Minkowski
spacetime. Consequently our cosmological model is one which
corresponds closely to a Robertson-Walker model insofar as the causal
structure of the spacetime and the large scale Hubble flow of the matter are
concerned. However, the local distribution of matter may be highly
inhomogeneous. Nevertheless, on scales small compared with those set by the
underlying Robertson-Walker model, Newtonian gravity holds to a very
good approximation. Apart from negligibly small regions of spacetime
which contain black holes or other strong field objects, I see no
reason to doubt that our universe is accurately described by this
model.

It should be noted that the metric (\ref{metric}) has many special
features, such as the presence of an irrotational, shear free
congruence---namely, the preferred observers of the underlying
Robertson-Walker metric---with respect to which the magnetic part of
the Weyl tensor vanishes. Ellis and van Elst \cite{ee} recently have
found that, apart from the Robertson-Walker models, there exist few,
if any, exact solutions of Einstein's with these special features.
However, there is no reason to expect these special features to survive when
higher order corrections to the metric are taken into account; indeed
the corrections mentioned above in footnote 1 already violate these
features. The non-existence of exact solutions with these features is
entirely consistent with the metric (\ref{metric}) providing an
excellent approximation to the actual spacetime metric and curvature of our
universe.

Consider, now, the propagation of an infinitesimal beam of photons.
All gravitational focusing and shearing effects are described by the
geodesic deviation equation
\begin{equation} 
\frac{d^2 \eta^a}{d \lambda^2} = - {R_{bcd}}^a k^b k^d \eta^c,
\label{gd}
\end{equation}
where $k^a$ is the tangent to the null geodesic, $\lambda$ is the
corresponding to affine parameter, and $\eta^a$ is the deviation
vector to an infinitesimally nearby null geodesic in the beam.  In a
Robertson-Walker model only Ricci curvature is present, and it causes
the beam to continually undergo a small degree of focusing. On the
other hand, in the cosmological model of Eq.~(\ref{metric}) in the
case where the matter is highly clumped, the Ricci tensor vanishes
along the geodesic, except for rare instances when the photon
propagates through a clump of matter. On these rare occasions, the
Ricci curvature briefly becomes extremely large compared with that of
the underlying Robertson-Walker model. The Weyl curvature also will be
small except in similarly rare instances of propagation through (or
very near) a sufficiently dense clump of matter. Thus, the local
history of a photon beam propagating in the spacetime of
Eq.~(\ref{metric}) can differ enormously from the local history of a
photon beam propagating in a Robertson-Walker model.

We wish to accurately calculate probability distributions for the net
result of the encounters of a photon beam with matter and Weyl
curvature in a spacetime of the form (\ref{metric}) under a wide range
of choices of underlying Robertson-Walker model and under a wide range
of assumptions about the inhomogeneities of the matter
distribution. To do so, we make use of the fact that the universe
appears to be Newtonian in a neighborhood of size $\ll R_H$ of any
Robertson-Walker observer. Thus, we can calculate a Newtonian
gravitational potential associated with the mass distribution in this
neighborhood\footnote{Justification for considering only the nearby
matter was given in section 1.3 of \cite{hw}.}, obtain its
corresponding spacetime curvature, and then propagate the photon beam
through this neighborhood using eq.(\ref{gd}). When the beam leaves
this neighborhood, we may view it as entering a similar Newtonian
neighborhood of another Robertson-Walker observer. Since the
Robertson-Walker observers are in relative motion, the photon beam
must be redshifted to describe it in the new Newtonian frame, and the
changes in the density of matter associated with the expansion of the universe
must be taken into account. However, apart from these effects of the
global cosmology, the propagation of the photon beam can be described
as a sequence of Newtonian encounters with the curvature resulting
from local mass distributions.

The above description of the propagation of photon beam provides
the basic rationale for our method for calculating probability
distributions for the amplification, shear, and rotation of images of
cosmologically distant sources.  We begin by choosing an underlying
Robertson-Walker model and specifying how matter is distributed in
galaxies (e.g., truncated isothermal balls of a given density and
radius). There is no restriction on how the galactic matter is
distributed other than the requirement that the average mass density
agree with that of the underlying Robertson-Walker model. We also may
allow changes in the physical structure of galaxies with time,
although we have not done so in most of our simulations.  Let $2 {\cal
R}$ denote the (comoving) average distance between galaxies.  We
perform a ``Monte Carlo'' propagation of a beam of photons {\it
backward} in time (starting from the present) in the following manner:
We imagine that our photon beam enters a ball of radius ${\cal R}$
centered on a galaxy with a random impact parameter.  We then
integrate Eq.~(\ref{gd}) through the ball using the curvature computed
from the Newtonian gravitational potential of the galaxy.  When the
photon beam exits from this ball, we use the underlying
Robertson-Walker model to update the frequency of the photon relative
to the local rest frame of the matter, and also to update the proper
radius corresponding to the comoving scale ${\cal R}$. Then we choose
another random impact parameter for entry of the photon into a new
ball and continue the integration of the geodesic deviation
equation. We then repeat this process until the photon has reached the
desired redshift. By re-doing this sequence of calculations a large
number of times, we build up good statistics on what happens to beams
of photons on our past light cone. From this we obtain---for any given
Robertson-Walker model and specification of the structure of
galaxies---good statistical information on the magnification, shear,
and rotation of images of (nearly) point sources at any redshift.

Note that in our Monte Carlo procedure, each photon in effect
``creates its own cosmological model'' during the course of its
propagation. Consequently, when multiple imaging occurs, our approach
does not provide a means of directly determining which images are
associated with the same source. However, in the case of point masses
or truncated isothermal balls, if a single lens dominates (as
plausibly would be the case in the strong lensing regime), then
standard analytic expressions can be used to obtain the the
relationship between the magnifications of primary and secondary
images of the same source.\footnote{A {\it primary image} is one
associated with a photon beam which has not undergone a caustic
whereas a {\it secondary image} is one associated with a beam that has
undergone a caustic, and, consequently, has entered the interior of
the observer's past. Every source has at least one primary image,
corresponding to a photon beam lying on the boundary of the past of
the observer. When the matter in galaxies is distributed in a
spherically symmetric manner, we have verified numerically \cite{hw}
that the total area carried by beams associated with primary images
very nearly equals the area of the boundary of the past of the
corresponding observer in the underlying Robertson-Walker model, at
least out to a redshift of $3$. This implies that essentially all
primary images lie on the boundary of the past of the observer, and,
thus, that essentially all sources have only one primary image.} This
allows us to statistically associate secondary images with primary
images.

As explained in \cite{hw}, our method can easily be adapted to take
account of sub-galactic structure on arbitrarily small scales, as
would be relevant for microlensing.\footnote{Clumping of matter on
small scales is relevant for lensing only if the angular size of the
Einstein radius of the lens is larger than the angular size of the
source. For the smallest sources of interest---namely quasars and
supernovae---this means that only clumping on mass scales larger than
$10^{-3} M_\odot$ is relevant.} However, in order to fully take into
account effects of the clustering of galaxies themselves, we would
need to choose our comoving scale ${\cal R}$ to be the scale of
clusters of galaxies, and we then would have to model the mass
distribution of the cluster. Fortunately, it does not appear necessary
that we take the clustering of galaxies into account in order to
obtain accurate results for the statistical distributions of the
magnification, shear, and rotation of images. To see this, consider
first the limit in which Ricci curvature dominates the lensing
effects, i.e., the galaxies are larger than their own Einstein radii
and the clustering of the galaxies also is not sufficient strong so as
to produce significant Weyl curvature.  Since, by Einstein's
equation, the Ricci curvature is determined by the matter distribution
in a completely local manner, the lensing effects of galaxies should
depend only very weakly on their clustering, since clustering should
merely produce some correlations in the times of passage of a photon
through different galaxies, but these effects should largely ``wash
out'' over cosmological distance scales. On the other hand, consider
the opposite extreme where galaxies lie well within their own Einstein
radii, and thus can be treated as ``point masses''. In this case, we
have shown \cite{hw} by a combination of analytic and numerical
arguments that arbitrary spherical clustering of the galaxies will
have at most a tiny effect on the lensing probability distributions
for the magnification, shear, and rotation of (nearly) point
sources. Although strong clustering of galaxies will create large scale
``cluster potentials'', the additional lensing effects produced by
these large scale potentials is almost exactly compensated by
``screening effects'' on the lensing by individual galaxies, and
almost no net change occurs in the probability distributions we
calculate. (On the other hand, clustering {\em would} still have an
important effect on some lensing quantities we do not calculate, such
as total bending angles.)  Thus, clustering of galaxies should be of
importance for the statistical lensing quantities we calculate only
when individual galaxies are larger than their own Einstein radii
(and thus individually contribute only Ricci curvature), but the
galaxies cluster into structures that produce significant Weyl
curvature. In these circumstances the neglect of the clustering of
galaxies should underestimate the lensing effects somewhat. However,
we do not believe that such circumstances arise frequently enough to
have an important influence on the statistical lensing quantities we
calculate.

Our approach is ideally suited to accurately calculate the effects of
many different, independent lenses on the image of a source. Although,
as noted above, in the strong lensing regime it is expected that a
single lens normally will dominate (for moderate source redshifts),
this need not be the case in the weak lensing regime. In particular,
our approach allows one to accurately calculate the probability that a
photon beam will encounter negligible curvature when traversing from
the source to the observer, and thus that the magnification of the
image of the source will be near its ``flat spacetime'' (i.e., ``empty
beam'') value. For reasonable matter distributions, this probability
is quite high for sources at redshifts $< 1$, so the ``empty beam''
formula for magnification generally provides at least as good an
approximation to the peak of the magnification probability
distribution as the Robertson-Walker (i.e., ``filled beam'') formula
\cite{hw}, \cite{h}. Consequently, the effects of lensing on the
analysis of data from type Ia supernovae can be quite significant
\cite{h}. Our method also has already been applied to calculate a
number of other effects of cosmological interest, specifically, the
correlations between magnification due to lensing and the number of
massive gas clouds through which the photon beam passes \cite{hw}, and
the probabilty for multiple imaging of very high redshift sources
\cite{hmq}. Many additional applications of our method are planned for
the near future. In particular, it is planned to use our method to
determine the source redshift range over which it may be assumed that
a single encounter with a lens dominates strong lensing effects, as
well as to accurately calculate the cumulative effects of the
sub-dominant lenses \cite{h2}.

\section*{Acknowledgements}

This research was supported by NSF grant PHY 95-14726 to the
University of Chicago.

\end{document}